\documentstyle[preprint,tighten,version2,aps]{revtex}
\def\goto{\mathop{\;\longrightarrow\;}}
{\makeatletter%
\gdef\labeleqs#1{{%
\edef\@currentlabel{%
\ifappendixon\appletter\fi
\ifsecnumbers\ifnum\c@secnum>0
\arabic{secnum}.\fi\fi\arabic{equation}}%
\label{#1}%
}}%
}
\begin{document}
\draft
\preprint{IFUP-TH 74/94}
\begin{title}
Master Wilson loop operators in large-$N$ lattice QCD$_2$.
\end{title}
\author{Paolo Rossi and Ettore Vicari}
\begin{instit}
Dipartimento di Fisica dell'Universit\`a and
I.N.F.N., I-56126 Pisa, Italy
\end{instit}
\begin{abstract}
An explicit solution is found for the most general independent correlation
functions in lattice QCD$_2$ with Wilson action.
The large-$N$ limit of these correlations may be used to reconstruct the
eigenvalue distributions
of Wilson loop operators for arbitrary loops. Properties of these spectral
densities are discussed in the region $\beta<\beta_c={1\over 2}$.
\end{abstract}


\narrowtext

In view of the renewed interest in the master field approach to large-$N$
matrix field theories~\cite{GG,D},
it may be convenient to improve our knowledge of those solvable
models that will eventually become suitable laboratories for testing
new methods and techniques.
Lattice  QCD$_2$ with free boundary conditions is such a solvable system.
In principle all correlation functions can be computed by a proper
gauge choice leading to a complete factorization of the functional
integral and by knowledge of the properties of invariant group integration.
The underlying single matrix problem has been solved a long time ago~\cite{GW},
and at first
glance not much can be added to what is already known.

However, as extensively discussed in Ref.~\cite{GG}, one of the problems that
must
be faced in even simpler models (like a model of $n$ independent Hermitian
matrices) is that of finding a master field description of correlations
involving different, and in general non-commuting, matrix field variables.
The purpose of the present letter is that of establishing a benchmark for this
problem,
in the form of an explicit large-$N$ expression for such correlations at
arbitrary
distance, and studying some properties of the eigenvalue distributions
of the (master) Wilson loop operators associated with these correlations.

Our starting point is the axial gauge formulation of lattice QCD$_2$
with Wilson's action presented in Refs.~\cite{BG,GW,R}. In this
formulation the expectation value of a Wilson loop is reduced to that of an
invariant product of field
variables in a one-dimensional principal chiral model with free boundary
conditions.
Let's  therefore focus on these models, defined by the lattice action
\begin{equation}
S\;=\; 2\beta N\sum_i {\rm Re}\, {\rm Tr}\, [ U_i U^\dagger_{i+1}]\;
\label{action}
\end{equation}
where $i$ are the lattice sites.
By  straightforward manipulations  one may show that the most general
nontrivial correlation one really needs to compute is
\begin{equation}
W_{l,k}\;=\; {1\over N} \langle {\rm Tr}\, (U_0 U_l^\dagger )^k \,\rangle
\label{wlk}
\end{equation}
where $l$ plays the role of the space distance and $k$ is a sort of ``winding
number''
for the loop.
In turn by the variable change $V_l=U_{l-1}U^\dagger_l$, the evaluation of
$W_{l,k}$ is reduced to computing the group integral appearing in the
relationship
\begin{equation}
W_{l,k}\;=\;{
\int dV_1...dV_l \;{1\over N} {\rm Tr}\,(V_1...V_l)^k \;\exp
(2N\beta \sum_{i=1}^l {\rm Re}\,{\rm Tr} V_i )
\over
\left[ \int dV \exp (2N\beta {\rm Re}\,{\rm Tr} V )\right]^l }\;.
\label{wlk2}
\end{equation}
This problem can be solved for arbitrary $N$ by a character expansion
(see e.g. \cite{DZ}):
\begin{equation}
\exp \left[ 2N\beta {\rm Re}\,{\rm Tr} \,V\,\right]\;\propto\;
\sum_{(r)} d_{(r)} {\tilde{z}}_{(r)}(\beta)\chi_{(r)}(V)\;,
\label{char}
\end{equation}
where $\chi_{(r)}$ are the characters and $d_{(r)}$ the dimensions of the
irreducible
 representations of $U(N)$.
Let's remind that the character coefficients ${\tilde{z}}_{(r)}(\beta)$ are
explicitly known
for all groups $U(N)$. As a consequence of Eqs.~(\ref{wlk2}-\ref{char})
we obtain
\begin{equation}
W_{l,k}\;=\; \int \prod_{i=1}^l \left[ dV_i \sum_{(r)}
d_{(r)} {\tilde{z}}_{(r)}(\beta)\chi_{(r)}(V_i)\right] {1\over N} {\rm
Tr}\,(V_1...V_l)^k
\;=\; {1\over N} \sum_{(r)} C_{k,(r)} d_{(r)} {\tilde{z}}_{(r)}^l\;,
\label{wlk3}
\end{equation}
where we have introduced the coefficients $C_{k,(r)}$ of
the decomposition
\begin{equation}
{\rm Tr}\, X^k\;=\; \sum_{(r)} C_{k,(r)} \;\chi_{(r)}(X)\;.
\label{ckr}
\end{equation}
It is now an exercise in representation theory to show that, for all $k<N$,
the only representations occurring in the r.h.s. of Eq.~(\ref{ckr})
are those in the form
\begin{equation}
\chi_{(j,k)}\equiv \chi_{(k-j+1,1^{j-1};0)}\;,
\label{chijk}
\end{equation}
and one may show that
\begin{equation}
{\rm Tr}\, X^k \;=\; \sum_{j=1}^k (-1)^{j+1} \chi_{(j,k)}(X)\;.
\label{exk}
\end{equation}
As a consequence we immediately obtain the final result holding for
$k<N$ in all $U(N)$ groups:
\begin{equation}
W_{l,k}\;=\; {1\over N} \sum_{j=1}^k (-1)^{j+1} d_{(j,k)} {\tilde{z}}_{(j,k)}^l
\;,
\label{wlk5}
\end{equation}
where we also know that
\begin{equation}
d_{(j,k)}\;=\; {(N+k-j)!\over (N-j)!} {(k-1)!\over k!(j-1)!(k-j)!} \;.
\label{djk}
\end{equation}

Further closed-form results may be obtained by restricting attention to large
$N$ and focusing
on the strong coupling domain. When $\beta< 1/2$ and $N$ is large~\cite{GS}
\begin{equation}
{\tilde{z}}_{(j,k)}\;=\; {N^k (N-j)!\over (N+k-j)!} \beta^k \left[ 1 + O\left(
\beta^{2N}\right)\right]
\;.
\label{tz}
\end{equation}
Under the same assumptions therefore
\begin{equation}
W_{l,k}\;=\; {N^{k-1}\over k} \sum_{j=1}^k {(-1)^{j+1}\over
(j-1)!(k-j)!} \left[ \prod_{i=0}^{k-1} \left( 1 + {k-j-i\over
N}\right)\right]^{1-l}
\beta^{kl}\;,
\label{wlk6}
\end{equation}
and one can finally show that
\begin{equation}
\lim_{N\rightarrow\infty}\;W_{l,k}\;=\; {(-1)^{k-1}\over k}
\left( \begin{array}{c} lk-2\\ k-1\end{array}\right)\beta^{kl}\;.
\label{wlk7}
\end{equation}
We have explicitly checked that Eq.~(\ref{wlk7}) satisfies the
Makeenko-Migdal equations~\cite{MM} for the large-$N$ limit of the
corresponding Wilson loops in
QCD$_2$.

Eq.~(\ref{wlk7}) is our main result, in that it is the quantity
that a master field approach to QCD$_2$ on the lattice should be able to
reproduce for arbitrary
$l$ and $k$ by substituting a single field configuration in the correlation
functions.

It is interesting to explore the properties of the Wilson loop
operator at a distance $l$, and in particular its eigenvalue distribution,
by constructing a generating function of the expectations $W_{l,k}$:
\begin{equation}
A_l(x)\;=\; \sum_{k=0}^\infty W_{l,k} x^k\;.
\label{al}
\end{equation}
{}From the Makeenko-Migdal equations it is possible to derive the
Schwinger-Dyson
equation satisfied by $A_l(x)$ in the strong coupling regime $\beta < 1/2$
and in the large-$N$ limit
\begin{equation}
\left( A_l(x)-1\right) A_l(x)^{l-1} \;=\; \beta^l x\;.
\label{MM}
\end{equation}
We checked explicitly the consistency of Eq.~(\ref{MM}) with our solutions
(\ref{wlk7}).
When $l=1$ one recognizes Gross-Witten's solution for the strong coupling phase
of the single plaquette models. When $l=2$ the solution
\begin{equation}
A_2(x)\;=\; {1\over 2}\left( \sqrt{ 1+4\beta^2 x} + 1\right)
\label{a2}
\end{equation}
is related to the generating function for the moments
of the energy density
\begin{equation}
{1\over N}\langle {\rm Tr} {1\over 1-\beta x (V_n+V_{n+1}^\dagger)}\rangle
\;=\; 1 + 2x\beta^2 + 2\sum_{k=1}^\infty (\beta x )^{2k} W_{2,k}
\;
=\; 2\beta^2 x + \sqrt{ 1 + 4 \beta^4 x^2}\;.
\label{poi}
\end{equation}
Incidentally, Eq.~(\ref{poi}) corrects a mistake in Ref.~\cite{BR}.

More generally, for arbitrary $l$, we may reexpress our results in terms of
the eigenvalue distribution $\rho_l(\theta)$ of the Wilson loop operator
$\prod_{i=1}^l V_i$, which is related to the generating function $A_l(x)$ by:
\begin{equation}
\rho_l(\theta)\;=\; {1\over \pi} \left[ {\rm Re} A_l (e^{i\theta})-{1\over 2}
\right] \;,
\label{rhol}
\end{equation}
and satisfing
\begin{equation}
\int_{-\pi}^\pi d\theta\,\rho_l(\theta)\;=\; 2W_{l,0}\,-\,1\;=\;1\;.
\label{norm}
\end{equation}

The study of the area dependence in the spectral density of continuum QCD$_2$
was pionered
a long time ago~\cite{R,DO} and recently reconsidered by many authors.
The lattice counterpart of this problem is the study of the dependence on $l$
of the functions $\rho_l(\theta)$, which can be performed numerically,
for arbitrary $l$, by summing up sufficiently many terms in the Fourier
expansion of
Eq.~(\ref{rhol}). In Figs.~\ref{fig1} and \ref{fig2} we plotted the spectral
densities as functions of $\theta$ for different values of $l$
at fixed values of $\beta$, and in particular when $\beta=\beta_c$.

As one may notice, the property
\begin{equation}
\rho_l(\theta)\goto_{l\rightarrow \infty}{1\over 2\pi}\;,
\label{limrho}
\end{equation}
which can be independently derived from the confinement properties of the
model~\cite{DO}, is exhibited by our solutions.  It is however quite
interesting to notice
that at $\beta=\beta_c$ and $\theta=\pi$ the equation satisfied by the
eigenvalue
density is
\begin{equation}
\left[ 1 - 2\pi \rho_l(\pi)\right] \left[
1+2\pi\rho_l(\pi)\right]^{l-1}\;=\;1\;.
\label{eqrhopi}
\end{equation}
Besides the strong coupling solution, which corresponds
to the values plotted in Fig.~\ref{fig2}, Eq.~(\ref{eqrhopi})
admits a second solution $\rho_l(\pi)=0$ (coinciding with the first one when
$l=1,2$).
This phenomenon leaves open the possibility of a compactified solution
in the weak coupling phase for arbitrary values of $l$.



\figure{
$\rho_l(\theta)$ at $\beta={1\over 3}$
and for various value of $l$. $\rho_l(-\theta)=\rho_l(\theta)$  by symmetry.
\label{fig1}}

\figure{
$\rho_l(\theta)$ at $\beta=\beta_c={1\over 2}$
and for various value of $l$.
\label{fig2}}

\end{document}